# Expanded flow rate range of high-resolution nanoDMAs via improved sample flow injection at the aerosol inlet slit.

Juan Fernandez de la Mora
Yale University, Mechanical Engineering Department, New Haven, CT 06520

**ABSTRACT**
High-resolution DMAs requiring hundreds of lit/min of sheath gas flow $Q$ to classify 1 nm particles have not been previously examined and optimized under the modest $Q$ values (tens of lit/min) needed to classify particles well above 10 nm. Here we study the resolving power $\mathcal{R}$ (based on the relative half width of the transfer function) of the Halfmini DMA, at sample flow rates $q>1$ lit/min. The charge-reduced electrosprayed ovalbumin protein used as test aerosol has an accurately determinable Gaussian mobility distribution, which limits the measurable $\mathcal{R}$ to at most 25-30. Non-ideal DMA response can however be precisely probed by comparing measured peak widths with the convolution of the protein's mobility distribution with the triangular Knutson-Whitby distribution associated to the finite $q_i/Q$ ratio. For the unmodified Halfmini DMA, $\mathcal{R}$ departs considerably from ideal at $q_i/Q$ values as modest as 2%, revealing a problem with the aerosol inlet flow. This imperfection is not removed by either increasing or decreasing the slit width, nor by reorienting axially the sample flow jet as it merges with the sheath gas. By introducing a ring with many perforations upstream of the inlet slit, which favors a more symmetric distribution of the sample flow over the slit perimeter, $\mathcal{R}$ values close to ideal are approached with $q_i/Q$ ratios as large as 6-12%. This improvement enables (in principle) classification of particles larger than 30 nm.

**Running title**: improved DMA inlet slit flow

## 1. Introduction

Differential mobility analyzers are capable of classifying an impressive range of particle sizes, from several hundred nm down to small sub-nanometer ions. For a cylindrical DMA of inner and outer electrode radii $R_1$, $R_2$ charged by a voltage difference $V$, and with classification length $L$, when the sample aerosol inlet and outlet flows $q_i$ and $q_o$ are matched ($q_i=q_o=q$), the selected *target* mobility $Z_V$ is related to the sheath gas flow rate $Q$ via (Knutson and Whitby, 1975)

$$Z_V = \frac{Q\ln(R_2/R_1)}{2\pi L V}. \tag{1}$$

While arbitrarily small $Z$ (large particle diameter) values may be classified at sufficiently small $Q$, a size limit is imposed by the need to achieve a certain resolving power, defined here as the inverse of the relative width of the mobility peak for a monomobile aerosol:

$$\mathcal{R}=V/\Delta V, \tag{2a}$$

Conventionally $V$ is the voltage at the maximum of the mobility peak and $\Delta V$ is the full peak width at half height. However, for relatively wide peaks where the position of the peak maximum is not as precisely defined as the two voltages $V^+$ and $V^-$ at half height, a definition based only on the pair of values $V^+$ and $V^-$ is preferable. One possibility is to substitute $V$ in (2a) either by the arithmetic or the geometric mean of $V^+$ and $V^-$. Here we adopt the alternative simpler but nonstandard definition based on the ratio of these voltages minus 1:

$$\mathcal{R}= V^+/V^- -1. \tag{2b}$$

Besides its simplicity, definition (2b) enjoys the practical advantage that the inferred resolving power is independent of whether the peak is represented in terms of mobility or the inverse mobility variable proportional to DMA voltage.

For a non-diffusing aerosol with matched inlet and outlet flow rates, $\mathcal{R}$ follows from the triangular transfer function of Knutson and Whitby (1975) as

$$\mathcal{R}^I=\frac{1+q/(2Q)}{1-q/(2Q)} -1, \tag{3}$$

which for small $q/Q$ is very close to $q/Q$. Accordingly, since a finite sample flow $q$ is required in order to obtain a measurable aerosol output, $Q$ must be at least $q\mathcal{R}$, resulting approximately in the following ideal minimal mobility for non-diffusing particles:

$$Z_{min}= \mathcal{R}q\frac{\ln(R_2/R_1)}{2\pi L V_{max}} \tag{4}$$

Table 1: Ideal upper size range for two DMAs, assuming $\mathcal{R}q$=20 Lit/min.

| DMA | $R_1$ (cm) | $R_2$ (cm) | $L$ (cm) | $V_{max}$ (kV) | $Z_{min}$ (cm$^2$/V/s) | $d_{max}$ (nm) |
|---|---|---|---|---|---|---|
| TSI 3071[a] | 0.937 | 1.958 | 44.44 | 11 | 7.99 10$^{-5}$ | 212 |
| Halfmini[b] | 0.4 | 0.7 | 2 | 5 | 0.00247 | 29.8 |
| Tapcom PDMA[c] | 1.8 | 2.4 | 6.5 | 10[d] | 0.000235 | 108 |

[a] Knutson and Whitby 1975. [b] Fernandez de la Mora and Kozlowski 2013. [c] Kallinger et al. 2011. [d] Value assumed for the purpose of this calculation

Table 1 compiles for reference the characteristics of three DMAs, with corresponding ideal largest classifiable particle diameters, assuming $\mathcal{R}q$=20 lit/min. The respective $d_{max}$ would be 346 nm, 43 nm and 165 nm for $\mathcal{R}q$=10 lit/min (163 nm 24 nm and 85 nm for $\mathcal{R}q$=30 lit/min). These ideal upper size limits are substantial for the 2 cm long Halfmini DMA, particularly considering that it is designed to handle laminarly sheath gas flow rates in excess of 800 lit/min in order to classify 1 nm particles with $\mathcal{R}$>40. However, one cannot take for granted that the assumed ideal behavior will be met in practice at $Q$ much smaller than the design values. The fluid dynamics of DMAs is in fact particularly delicate and prone to instabilities in the region where the sample gas enters the classification region, where a mixing layer forms between the aerosol and sheath gas streams, generally having different speeds. Although such mixing layers are always unstable under inviscid conditions, they may be expected to be stable at sufficiently small Reynolds numbers for the annular aerosol jet, defined here as

$$Re_{slit}=q_i/(2\pi R_2 \nu), \qquad (5)$$

where $\nu$ is the kinematic viscosity of the sample gas. This expectation is met in reality by the Halfmini DMA at small enough $q_i$, where high resolving powers are achieved even though the sheath gas velocity is much larger than the aerosol velocity. However, as shown in Figure 1a, the resolution of this DMA drops substantially below the ideal value (see section 2) at increasing $q/Q$. As a result, the upper particle diameter limit compatible with a given resolving power is considerably smaller than the ideal values listed in Table 1. This non-ideality has other undesirable consequences, even when analyzing 1-5 nm particles. For example, due to high diffusion losses in this size range, recent tandem DMA studies of clusters have required relatively large sample flow rates (Attoui et al., 2013a, b), which reduce the resolution, limiting the ability to distinguish clusters containing $n$ and $n+1$ molecules with relatively large $n$. For these and other related reasons, the present work investigates the source of non-idealities arising at the aerosol inlet slit of the Halfmini DMA, as well as several approaches to moderate or cancel them.

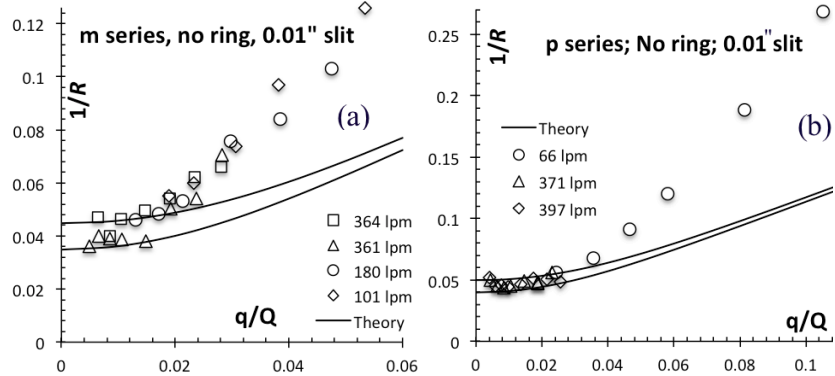

Figure 1: Peak width for an ovalbumin aerosol analyzed by two Halfmini DMAs, both showing large departures at $q/Q$>2% from the ideal behavior (accounting for the finite width of the test aerosol; Section 2). (a) earlier DMA model (*m series*) with radially pointing aerosol jet. (b) New DMA (*p series*) with axially pointing aerosol jet

A major source of non-ideality will be seen to be associated to lack of axisymmetry of the aerosol stream as it enters the analyzing region. The average flow rate per unit slit length is $q'=q/(2\pi R_2)$, but asymmetries make $q'$ dependent on the angular position $\theta$ along the slit perimeter, resulting in local *q'* values considerably larger than the average, especially in the slit region near the inlet tube bringing the aerosol sample. This asymmetry naturally results in a *θ*-dependent width of the DMA response function. It also increases the local Reynolds number of the entering aerosol jet, promoting mixing layer instabilities at unnaturally low *q* values. The aerosol inlet asymmetry has been considered in the past, but we are not aware of a satisfactory solution. TSI's widely used 3071 DMA includes means evidently intended to produce a relatively axially symmetric sheath gas stream, but has no obvious analogous elements to provide an axisymmetric aerosol flow. For undisclosed reasons, the Vienna DMA (Reischl et al., 1991) introduces the sample aerosol tangentially. According to Rosell et al. (1996) the resulting swirl "*... plays the positive role of transporting the aerosol into the full circumference of the DMA, though at the cost of inducing certain instabilities in the main analyzer flow*". These *instabilities* took the form of a doubly humped transfer function. Filling much of the annular space upstream the inlet slit with a ring resulted in a single peak, apparently by dampening the swirl, though "*at the cost of a non-uniform entering aerosol flow, which increases locally the ratio $q_i/Q$, thus decreasing resolution.*" This precedent led Eichler (1997) to design the annular inlet chamber of his own DMA such that the resistance for the gas to flow along the circumference of a relatively wide annulus would be much smaller than that required to cross the relatively narrow slit. Unfortunately this design was not tested at the small $q/Q$ ratios relevant to classify large particles. A limited number of unpublished subsequent tests in the Herrmann and the Halfmini DMAs explored the effect of widening or narrowing the inlet slit, but were not successful at removing the unnaturally large resolution loss observed at increasing *q*. Since the pressure drop across the slit increases quadratically with the inverse slit width, these failures suggested that Eichler's strategy might not suffice to solve the problem. However, a more systematic investigation was not pursued. Our current goal is accordingly to fill this gap.

**2. Theoretical loss of resolution due to the lack of a monomobile size standard**
For an aerosol with a Dirac distribution of mobilities centered at *Z*, the ideal DMA output aerosol concentration at a *target* mobility $Z_V$ [related to the DMA control variables via equation (1)] is:

$T(Z_V, Z)=0$ when $|Z/Z_V-1|>q'$ (6a)

$T(Z_V, Z)=1-|Z/Z_V-1|/q'$ when $0<|Z/Z_V-1|<q'$, (6a)

with $q'=q/Q$. (7)

For an aerosol with a mobility distribution *f(Z)*, the ideal aerosol output concentration $F(Z_V)$ for a DMA setting $Z_V$ includes contributions from a range in input *Z* values, given by superposition as:

$$F(Z_V)=\int f(Z)\, T(Z_V, Z)\, dZ= \int_{Z_V(1-q')}^{Z_V(1+q')} f(Z)(1-\frac{|Z/Z_V-1|}{q'})dZ.$$ (8)

As we shall see, the mobility distribution of the ovalbumin aerosol primarily used in this work is well approximated by a Gaussian,

$$f(Z)=\sqrt{\frac{a}{\pi}}e^{-a(Z-Z_o)^2},$$ (9)

for which the convolution integral (8) may be written analytically as (10b), in terms of the dimensionless variables (10a):

$$x=Z_o/Z_V,\ y=aZ_o^2 \text{ and } q' \tag{10a}$$

$$F(x)=x\frac{e^{-y(1-\frac{1+q'}{x})^2}-2e^{-y(1-\frac{1}{x})^2}+e^{-y(1-\frac{1-q'}{x})^2}}{2q'\sqrt{\pi y}}+\frac{1-x}{q'}Erf(\frac{x-1}{y^{-1/2}x})+\frac{x-1-q'}{2q'}Erf(\frac{x-1-q'}{y^{-1/2}x})+\frac{x+q-1}{2q'}Erf(\frac{x+q'-1}{y^{-1/2}x}). \tag{10b}$$

The second group $y=aZ_o^2$ is a measure of the intrinsic mobility width of the protein sample used, simply associated to the maximum resolving power $\mathcal{R}_o$ that can be measured with this imperfect aerosol standard:

$$\mathcal{R}_o=\sqrt{\frac{4ln(2)}{aZ_o^2}} \tag{11}$$

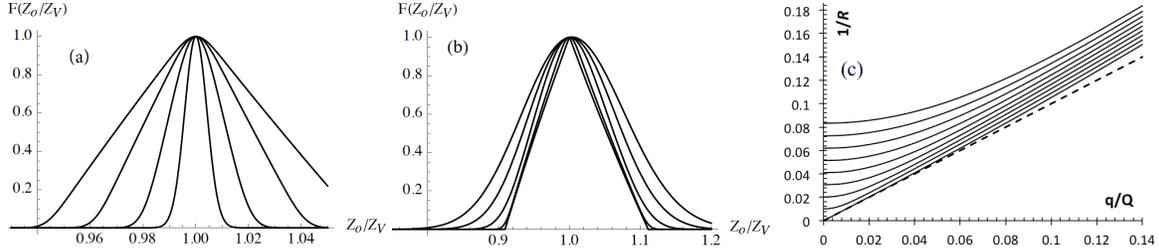

Figure 2: (a, b) Normalized peak shapes predicted by Equation (10b) for various ($\mathcal{R}_o$, $q'$) pairs. (a): $\mathcal{R}_o$=100, $q'$=10$^{-7}$, 0.02, 0.04, 0.06 (center to sides). (b): $q'$=0.1, $\mathcal{R}_o$=1899, 60, 19, 10.4, 7.35 (center to sides). (c) $\mathcal{R}_a(\mathcal{R}_o, q')$ curves from Equation (10) for $1/\mathcal{R}_o$=0; 0.01; 0.02; 0.03; 0.04; 0.05; 0.06; 0.07; 0.08 (bottom to top). The straight dotted line is $\mathcal{R}_a$=q/Q.

Several instances of normalized peak shapes are given in Figures 2a-b for various values of the parameters $\mathcal{R}_o$ and $q'$. At $q'$=0.1, in (b), for large $\mathcal{R}_o$ (narrow mobility standard) the curves are close to the ideal triangle. At $\mathcal{R}_o$=60 the triangle is distorted primarily at its vertices. The inverse of the half width $\Delta Z_V$ of these distribution normalized by $Z_o$ is an apparent resolving power, to be referred to as $\mathcal{R}_a$. $\mathcal{R}_a$ is a function of the two dimensionless groups $\mathcal{R}_o$ and $q'$ represented in Figure 2c versus $q'$ for selected $\mathcal{R}_o$ values:

$$\mathcal{R}_a = \mathcal{R}_a(\mathcal{R}_o, q'). \tag{12}$$

The indicated values of the parameter $\mathcal{R}_o$ associated with each line may also be read approximately from the graph as the limit of the function $\mathcal{R}_a(\mathcal{R}_o, q')$ as $q'\to 0$. The predicted curves are approximately horizontal at $q'$=0, with an apparent resolving power determined by the quality of the mobility standard, with little $q'$ dependence. Figure 2 shows that the ideal Knutson-Whitby curve (3) is substantially modified by the use of an imperfect mobility standard. For instance, for $1/\mathcal{R}_o$=0.04, at $q'$=0.04, $1/\mathcal{R}_o$ is displaced from 4% to 5.71%. Figure 2c includes two lines going through the origin (corresponding to $\mathcal{R}_o$=0). The straight dotted line is $\mathcal{R}_a^{-1}=q'$, corresponding to the conventional definition of the resolving power, while the curved continuous line is equation (3). This apparently small difference does in fact reduce noticeably the differences to be observed between theory and experiments.

While Equation (10) applies approximately to our weakly-diffusing protein particles having a finite mobility distribution, one may argue that it applies equally to our alternative standard of diffusing monomobile particles (C$_{12}$) by simply substituting the diffusive broadening (also Gaussian; Stolzenburg, 1988; Tammet, 1970; Rosell et al., 1996) in lieu of the intrinsic width. The main practical difference between both situations is that the intrinsic width of the protein standard is independent of $Q$ (weak diffusive broadening) while the diffusive peak width of our ion standard scales with $Q^{-1/2}$.

## 3. Experimental
**General considerations for generating size standards above 5 nm**.
A key obstacle complicating the precise determination of DMA performance in the size range of tens of nm has been the lack of methods to produce highly monomobile aerosols in the size range above 5 nm. Prior to the development of electrospray ionization (Fenn et al, 1989), the conventional means to produce a test aerosol to characterize one DMA was size classification of a polydisperse aerosol with another DMA (ideally identical to the first DMA; Rader and McMurry, 1986), combined with a deconvolution process to infer the individual DMA transfer functions from the measured transfer function of both DMAs in tandem (TDMA). Here we follow a hybrid of the electrospray and the TDMA methods. We use electrosprayed protein ions as the test aerosol, which is, however, not exactly monomobile due to the coexistence of a range of protein structures. The associated $\mathcal{R}$ for an electrosprayed protein may nonetheless be as large as 25-30 (Fernandez de la Mora, 2015). Unlike the output of the first DMA in the TDMA method, the intrinsic mobility distribution of the protein standard may be measured accurately. Indeed, proteins are small enough to be classifiable with high-resolution DMAs previously characterized with exactly monomobile molecular standards. Classification of a protein aerosol does not generally narrow down its size distribution, because the process of interconversion of one protein structure into another occurs in times short compared to the classification time. Accordingly, the protein standard method relies on one DMA alone, and is free from the problematic TDMA hypothesis that two different DMAs have identical responses. Some form of inversion is still necessary to extract DMA transfer function information from its measured convolution with the natural width of the protein standard. However, an inversion process is not essential to identify and correct DMA imperfections. We will simply use the results of section 2 on the theoretical response of the ideal triangular Knutson-Whitby transfer function combined with the measured protein size distribution, to infer unambiguously the ideal response of the DMA to our imperfect protein aerosol. Comparison of this theoretical ideal response to the real measured response directly reveals whether or not flow problems exist (as in Figure 1), and suggests schemes to correct them.

**Materials**
Two test aerosols generated by positive mode electrosprays were used. One involved the dimer ion of tetradodecylammonium bromide ($C_{12}$; Fluka) dissolved at about 1 mM in ethanol. The other relied on charge-reduced Ovalbumin ions. A Stock solution 90 µM of ovalbumin (Sigma, lyophilized, lot #SLBK1399) and used without further purification) was first prepared in deionized water (no buffer) and kept in a refrigerator. 1.8 µM protein solutions were freshly prepared before each set of measurements by adding 2 µL of the stock solution to 100 µl of one of two buffers: either 100 mM aqueous triethylammonium formate (TEAF) or 100 mM aqueous ammonium acetate. Charge-reduced mobility spectra for the buffers were taken right before adding the protein, to verify that they were completely free from involatile impurities that would form particle residues. The just prepared protein solutions were equally seen to be free from residues, but slowly evolved producing a high-mobility peak whose height increased over a time scale of 1 hour. The two buffers led to similar rates of protein degradation.

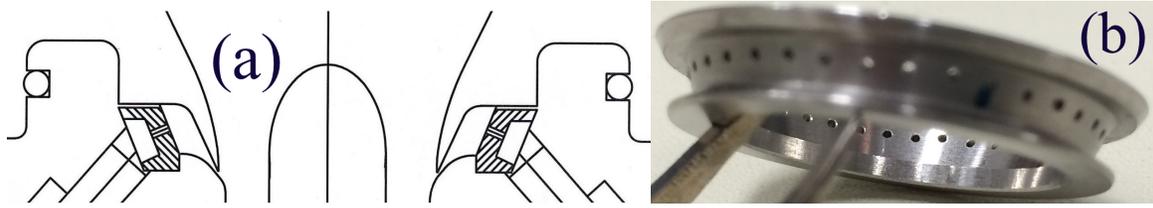

Figure 3: Elements controlling the flow pattern at the aerosol inlet slit. (a) Side view sketch of the new annular chamber preceding the inlet slit. Notice the curved and down-aiming lower lip in the slit, and the perforated ring (ruled) interposed between the aerosol inlet tube and the slit. (b) Image of the 24-hole ring with 0.02" perforations.

**DMA**. Key characteristics of the cylindrical Halfmini DMA used are listed in Table 1. Several geometrical variations were tested. In the original version (*m*-series; Fernandez de la Mora & Kozlowski, 2013) the sample aerosol jet projecting into the analyzing region at the inlet slit has a substantial initial radial velocity. The sample jet might then conceivably separate from (and then reattach to) the outer DMA electrode, creating a recirculation bubble that could periodically shed vortices at large enough $q$. In order to moderate this risk, a new DMA geometry (*p*-series) was manufactured where, not only the upper lip of the inlet slit, but also its lower lip are directed substantially downwards. In addition, a relatively large radius of curvature is introduced on the lower slit lip following the exit channel, such that the aerosol jet curves gently into the axial direction, reducing its tendency to separate (Figure 3a). Tests of this slit configuration showed slight performance improvements at moderate $q/Q$ (Figure 1b), but were still far from ideal. In order to correct for possible lack of axisymmetry in the aerosol jet entering the analyzing region, a substantial restriction in the path of the aerosol inlet flow was introduced between the inlet slit and the end of the 3 mm ID tube bringing the aerosol from outside the DMA (lower right and left corners of Figure 3a). This restriction is a ring with a number $N$ of symmetrically arranged holes (Figure 3a,b). This perforated ring concept was previously applied to a completely new though still untested DMA design aimed at classifying 60 nm particles (Perez-Lorenzo and Fernandez de la Mora 2017), and owes much to Mr. Luis J. Perez Lorenzo. Three rings were tested, with $N$ values of 24, 36 and 72. The hole diameters $D$ on the first two rings were 0.020" (0.5 mm), all aligned in a single row in the middle of the ring (figure 3b). The 72-hole ring had three rows of 24 holes, with all diameters of 0.020" in one row, all diameters of 0.028" in another row, and a third row with a mix of fourteen 0.020" holes, eight 0.028 holes and two holes blocked by broken drills. The individual jets formed by each of these orifices have substantial Reynolds numbers. For instance, at $q$=10 lit/min, $D$=0.02" $Re_o=4q/(N\pi D\nu)$ = 1180 for $N$=24 and $Re_o$=787 for $N$= 36. These jets surely become turbulent even at $q$ below 10 lit/min. Much of the jet momentum and turbulence, however, is damped and spread laterally as the jet collides with the wall of the upper lip and flows towards the slit exit (Figure 3a). For the hybrid 72 hole ring, assuming that the exit velocity (driven by a fixed pressure drop) is the same for both hole diameters and equal to $q/(N_1\pi D_1^2/4+ N_2\pi D_2^2/4)$=81.6 cm/s at $q$=10 lit/min, $Re_o$ = 381 and 272 for the 0.028" and 0.02" orifices. Besides the cylindrical bullet-shaped inner DMA electrode (Reischl, 1991) characterized in Table 1, a slightly conical inner electrode with a 2º half angle, reaching a radius of 4 mm at the outlet slit, was tested with the 24 and 36-hole rings. This shape is meant to accelerate the flow in the analyzing region, stabilizing the boundary layer at arbitrarily large $Q$ (Hoppel, 1970; Fernandez de la Mora 2002).

Another potentially important variable is the height $\delta$ of the inlet slit, which determines the velocity of the annular aerosol stream as it mixes with the sheath gas, $U=q/(2\pi R_2 \delta \sin 30^o)$. $\delta$ can be controlled by adding shim stock pieces between the DMA parts containing the upper and lower lips of the inlet slit, without affecting the centering or the leak-tightness of the instrument. Three configurations were tested with $\delta$= 0.010", 0.015" and 0.020".

The sample inlet flow rate $q_i$ varied from 1.2 to 10 Lit/min. It was controlled and measured by a valve and a ball flowmeter that was calibrated in that full flow rate range. The sample gas (filtered and dried laboratory compressed air) passes through an electrospray chamber, and entrains the electrosprayed species into a charge-reduction chamber (Kaufman et al. 1996) containing a nickel ring coated with 10 mCi of the beta emitter Ni-63. The interior wall of the Ni-63 coated ring is 1" long and 1" in diameter. The electrospraying chamber is separated from the charge-reducing chamber by a 50% transparent metal screen meant to avoid penetration of negative ions from the charge reduction chamber into the electrospray chamber (Fernandez de la Mora 2015). The charge-reduced electrospray species created in this chamber then enter into the polydisperse aerosol inlet to the DMA. The DMA-selected monodisperse aerosol is captured by a fast (<100 ms response time) and sensitive (0.12 fA noise) Faraday cup electrometer commercialized by SEADM (Fernandez de la Mora et al, 2016), which measures the current $I$ carried by the sampled gas. Its flow rate $q_o$ is not controlled, but is automatically equal to $q_i$ because the sheath gas flow circuit operates in closed loop, with a refrigeration system and a HEPA filter following the blower. The sheath gas flow rate $Q$ was controlled by fixing the speed of rotation of the blower, but was not directly measured. Instead, $Q$ is determined from the voltage $V$ at the center of the most mobile peak appearing in a $I(V_{DMA})$ spectrum, according to the relation (1). $Z$ is the electrical mobility under ambient conditions of either the $C_{12}^+$ ($Z_{C12}$=0.714 and 0.493 cm$^2$/V/s for the monomer and dimer ions, according to Ude and Fernandez de la Mora, 2005) or the Ovalbumin$^+$ ion ($Z_{Ova}$=0.0671 cm$^2$/V/s, according Fernandez de la Mora, 2015) used as mobility standards. The ratio $ZV/Q$ for the DMA with the conical inner electrode was found to be 1/0.866 times larger than for the cylindrical bullet by operating the two DMAs under identical flow conditions and recording the ratio of voltages ($V_{cyl}/V_{cone}$=0.866) at which the ovalbumin ion appeared.

Raw mobility spectra were obtained by fixing all flow rates and the setting of the electrospray source, sweeping the voltage $V_{DMA}$ of the inner electrode while grounding the outer electrode, and recording (in a computer, with the help of a data acquisition card) the aerosol current collected as a function of $V_{DMA}$. The high voltage power supply used was from Applied Kilovolts (±10 kV, 30 ms response time from +10 to -10 kV).

## 3. Results
### 3.1 Effect of the new downward aerosol entry geometry.
Figure 1 compares peak widths in the earlier *m series* inlet configuration and the new *p series* inlet configuration (depicted in Figure 2), neither including a perforated ring. One can see a slight resolution increase in the new design (b), but the strong non-ideality already discussed in relation to Figure 1a survives in Figure 1b. No improvements were observed either with the conical inner electrode, which provides additional stabilization in the aerosol inlet region. We conclude that the main inlet problem is not due to the radial component of the inlet aerosol jet entering the separation region in the *m series*.

**3.2 Effect of the perforated inlet ring.**
Figure 4 shows DMA resolution versus $q/Q$ measurements for two DMA configurations including perforated rings, with all the data sets but one relying on the dimer ion of $C_{12}$. Before discussing the implications of these data on DMA performance, let us first consider the insights they provide on the test aerosols. The $C_{12}$ dimer is strictly monodisperse and produces narrow peaks at small $q$ and large $Q$. However, these peaks are substantially widened by Brownian diffusion, whence the various curves shown move up at decreasing $Q$. This results in less discriminating resolution information (smaller $\mathcal{R}_o$) than obtainable with the protein standard at sheath gas flow rates below about 140 L/min. Figure 4a includes one curve for the ovalbumin standard, for which $\mathcal{R}_o \sim 22$ (26 in one data set in Figure 1, relying on a freshly made solution). The data at 600 lit/min demonstrate resolving powers in excess of 35 at small $q'$. Using monomobile ions larger than $C_{12}$ but smaller than ovalbumin we have recently demonstrated resolving powers of this DMA approaching 50 (Barrios and Fernandez de la Mora 2016). Therefore, the relative peak width of 3.5-4.5% found here for ovalbumin is not limited by imperfections of the DMA, but is rather the intrinsic width of the protein aerosol.

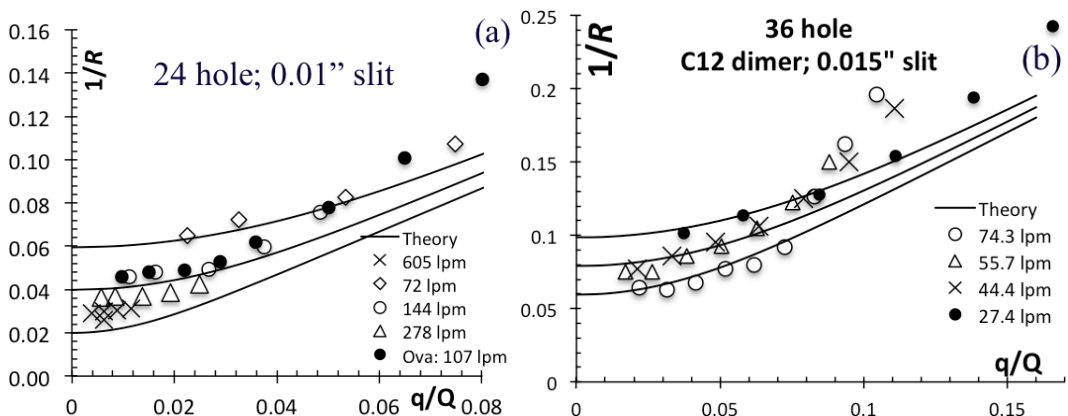

Figure 4: Peak widths with 24 and 36 hole rings installed in the DMA. (a) higher $Q$ values with $C_{12}$ dimer and ovalbumin standards. (b) Lower $Q$ values with $C_{12}$ dimer standard.

We now focus on comparing the theoretical curves with the data of Figure 4, obtained in DMAs with two different ring configurations. As in Figure 1 the response is almost ideal up to a critical value $q'^*$ of $q/Q$, with substantial departures from theory at $q/Q > q'^*$. There is, however, a drastic beneficial effect of the rings in that $q'^*$ is greatly increased. For instance, $q'^*$ is now 4% and 6% at 144 and 72 L/min (Figure 4a), versus ~2% in the ring-less configurations at 100 L/min. Notice also a substantial increase of $q'^*$ with decreasing $Q$, a point documented in Figure 5 for the data of Figure 4 and for others to be later discussed. These experimental $q'^*(Q)$ data display substantial scatter, in part due to the sparse grid of $q$ values used and the lack of a sharply defined (critical) onset of non-ideality, but also due to variations in the quality of the protein standard as well as errors in the determination of the peak width. It is nonetheless clear that an almost ideal response is available over a broad range of $q$, $Q$ values, including the conditions quoted in Table 1 enabling classification of 30 nm diameter particles ($q$=1 L/min; $Q$=20 L/min).

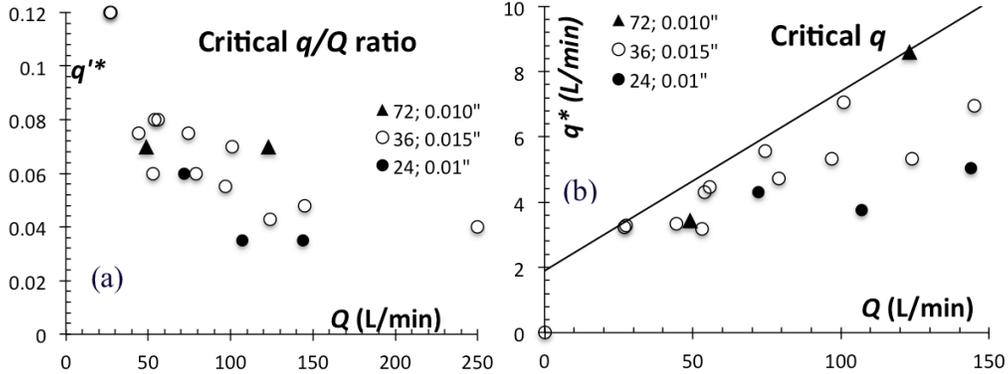

Figure 5: (a) Sheath gas flow rate dependence of the critical ratio $q'^*$ of $q/Q$ above which the DMA response ceases to be ideal. $N$; $\delta''$ in the legend refer to the number of ring holes and the slit width. (b) Same information as in (a) in terms of the critical sample flow rate, with stable conditions lying below the continuous line.

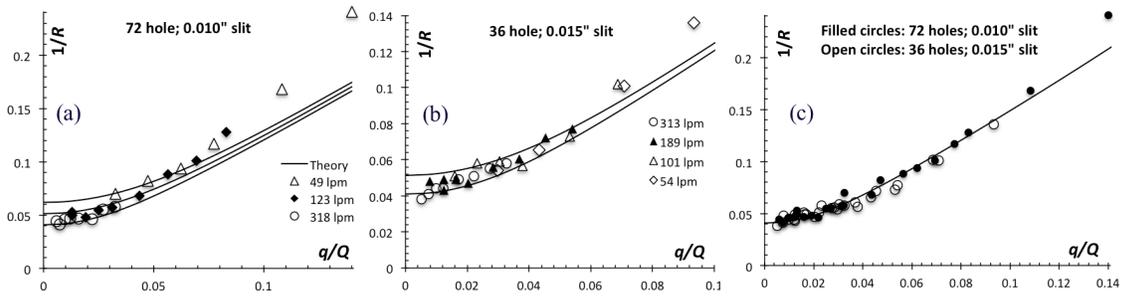

Figure 6: DMA performance with the 72 (a) and the 36 (b) hole rings at various flow rates of sheath gas. (c) Direct comparison of (a) and (b) without distinction of flow rates, showing a minimal effect of the number of holes. The theory line in (c) is artificially merged with the data by plotting $\mathcal{R}_a^{-1}$ versus $0.8q/Q$.

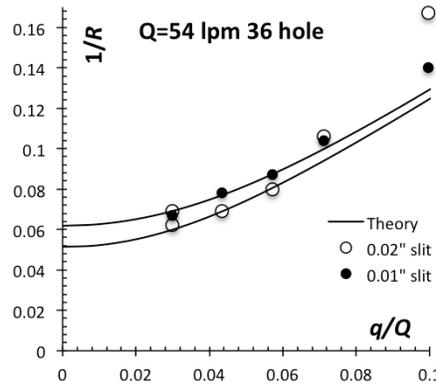

Figure 7: Effect of the inlet slit width at $Q$=54 L/min, with a wider ideal domain for the narrower slit at the largest $q/Q$.

Peak widths for other ring and slit configurations are shown in Figures 6 and 7. Figure 6 shows relatively little differences between either the 72 and 36 hole rings, or the 0.010" and the 0.015" slits, similarly as seen in Figure 4 for these two slit widths and for the 24 and 36 hole rings. The data of Figure 6, however, are more discriminating than those of Figure 4 thanks to the use of fresher ovalbumin solutions. There is no noticeable departure from the theoretical curves until $q/Q$ reaches 6-7%, which exceeds the range of interest for applications requiring $\mathcal{R}$ >15. Even in the region where non-idealities are present, they are relatively mild, and are absorbed (except for the last datum at $q/Q$=0.14) when the $q/Q$ scale is shrunk by 20% (continuous line in Figure 7c).

### 3.3 Effect of the slit width $\delta$.

Very little influence of increasing the slit width $\delta$ from 0.010 to 0.015" has been seen in the tests already described, which however included also different ring configurations. We have therefore run tests varying $\delta$ more broadly, from 0.010" to 0.020", under otherwise identical conditions (Figure 7). The main difference seen at moderate $q/Q$ is associated to a slight degradation of the protein sample during the first measurement series ($\delta$=0.020"). This irrelevant difference aside, both slits yield an almost ideal response up $q/Q$~6%. Beyond this value, the wider slit exhibits stronger non-idealities with a slightly smaller $q^{'*}$ onset. This observation is puzzling. Since $Re_{slit}$ in Equation (5) is independent of $\delta$, no $\delta$ effect would be expected if the source of non-ideality was instability of the mixing layer above a critical value of $Re_{slit}$. Alternatively, from the strictly inviscid point of view, as $q$ increases, an increased instability would be expected once the gas velocity at the slit exceeds the sheath gas velocity. But this effect would be delayed to larger $q$ in the wider slit, contrary to what is observed. One imaginable advantage associated to the narrower slit is its increased effectiveness to weaken flow turbulence surviving in the jets produced by the holes in the ring. However, these jets are substantially weakened in the 72-hole ring, without resulting in a noticeable narrowing of the peaks. Understanding and moderating the weaker non-idealities remaining after the introduction of the rings is of great practical interest, but was not pursued further due to the complexity of the various phenomena involved.

### 3.4 Peak shape evolution at increasing $q/Q$

In an effort to identify the nature of the flow problem resulting in performance degradation at $q/Q>q^{'*}(Q)$, we have compared the observed and predicted peak shapes, which evidently contain more information than just their width. This comparison was initially motivated by the observation of non-Gaussian low-mobility tails in the peaks at increasing $q/Q$, which we thought might be associated to unsteady vortex shedding from mixing layer instabilities. Figure 8 includes the full series of peak shapes for some of the spectra whose widths are reported in figure 7. The data are given as points, the gray curves are Gaussian fittings, and the continuous black lines are fittings to the convolution (10) represented in terms of the inverse quantity $Z_o/Z_V$, proportional to DMA voltage. In spectra (a)-(c) the peak shape is primarily due to the intrinsic protein mobility distribution, so the convolution shape is close to Gaussian. These relatively narrow peaks reveal the presence of a well-resolved high mobility contaminant at the left, and a partially unresolved low-mobility contaminant to the right. Spectrum (d), with $q/Q$=7.1%, is already noticeably non-Gaussian, with a maximum clearly displaced to the left of the Gaussian peak, and with its high mobility tail on the left falling faster than the right tail. Our original interpretation of the low mobility tail as due to mixing layer instabilities is accordingly incorrect, as this tail is very close to the predicted asymmetric peak shape. Figures 8e and 8f correspond to the maximal sample flow rate in Figure 7, both with $q/Q=0.1$. Although spectrum (f) associated to the wider slit is clearly wider than (e) its shape still conforms well to the theoretical shape, though with an artificially large $q'$ value of 0.13 (versus an experimental value of 0.1).

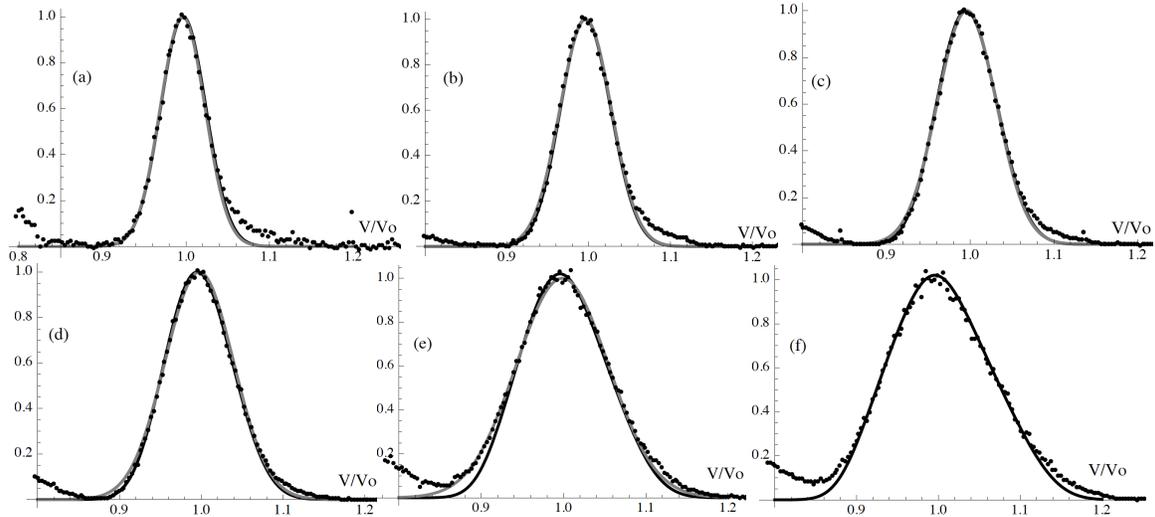

Figure 8: Normalized protein spectra from some data of Figure 7. (a-e): $\delta=0.010"$ at $q/Q$ increasing from (a) to (e). (f): $\delta=0.020"$, $q/Q=0.1$. The gray and black lines are fittings to a Gaussian and to the convolution curve (10), respectively.

The trend towards widening the $q/Q$ range yielding an ideal DMA response continues at decreasing $Q$, to the point that no disagreement with theory is observable at $Q=27$ L/min, not only for the peak width (Figure 9a), but also for its shape (Figure 9b).

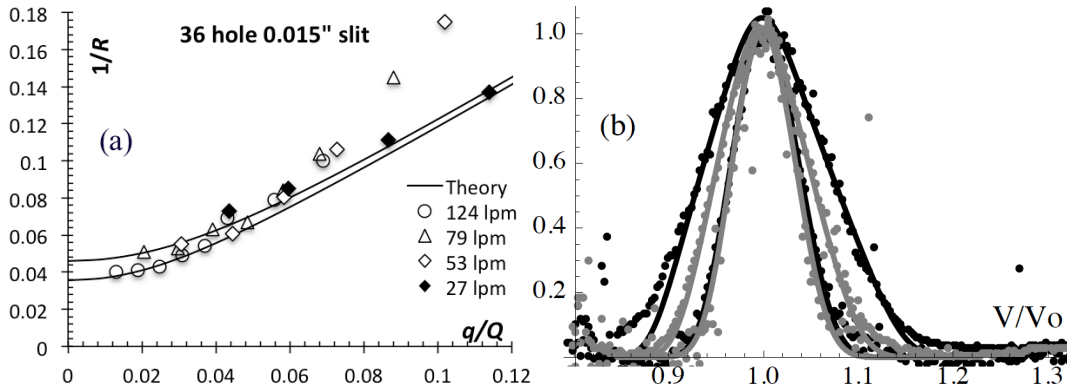

Figure 9: DMA response at modest sheath gas flow rates, with essentially ideal response at $Q=27$ L/min at all sample flows investigated. (a) Peak widths. (b) Peak structures for the series at $Q=27$ L/min (gray, black, gray, black at increasing $q$)

## 4. Discussion
### Coverage of relevant operating conditions
Brownian motion, the finite size distribution of the protein aerosol and the finite $q/Q$ ratio are fundamental sources of DMA transfer function broadening. However, the influence of diffusion is minimal for ovalbumin, while the other two effects are included here quantitatively. Additional broadening effects not easily quantifiable will be termed *non-idealities*. They include geometric imperfections (the finite width of the inlet and outlet slit, lack of axisymmetry in the electrodes, etc.) as well as flow imperfections (unsteadiness or turbulence, lack of axisymmetry in the aerosol inlet flow, etc.).

Since measured $\mathcal{R}$ values near 50 have been obtained with large ions at select ($q$, $Q$) values, geometric imperfections may be ignored at the scale of the maximum resolving power $\mathcal{R}_o$=25-28 measurable with our protein aerosol. Therefore, the $\mathcal{R}_o$ value inferred for the protein aerosol at small $q/Q$ is in fact the actual intrinsic width of the protein standard. Consequently, all expected non-idealities are accounted for in the theoretical curves except for those associated to (i) flow unsteadiness, (ii) lack of axisymmetry in the sample aerosol. For a given DMA, working gas and thermodynamic conditions, any existing flow imperfection will be governed by the geometry of the sample inlet and the pair of flow parameters $q$ and $Q$. Given that the various rings and slits tested have resulted in ideal behavior over a range of favorable flow conditions, it is safe to assume that the perforated ring inlets result in axisymmetric aerosol flow injection. Non-idealities observed under unfavorable conditions of ($q$, $Q$) must therefore be due to the onset of flow instabilities. For given ring and slit geometries, these instabilities may depend only on the two Reynolds numbers associated to the two streams, governed strictly by the pair of numbers ($q$, $Q$). A fairly broad range of ($q$, $Q$) pairs has ben investigated, as summarized in Figure 5. This incomplete coverage of parameter space has shown that, for $q/Q \leq 5\%$, ideal behavior is attained as long as $Q$<124L/min. This permits classifying 30 nm particles with a resolving power of 20. Flow rates smaller than 27 L/min have not been investigated, but the trend found favoring ideal behavior at decreasing $Q$ suggests a high probability that the flow will be as stable at 10 and 20 L/min as found at 27 L/min, offering a size range substantially above 30 nm to those willing to compromise the resolving power.

**Choice of inlet geometry**

Although favorable operating conditions have been found for all combinations of slit widths and number of ring orifices, there appears to be a slight advantage to the narrowest slit $\delta$=0.010" and for either the 36 and 72-hole ring over the 24 orifice ring (Figure 5). Given the difficulty and expense of making so many small perforations, a 36-hole ring seems adequate for future versions of this DMA.

**Leading causes of non-ideality**

The radical advantage found in association to the perforated rings tested in this work confirms that a major problem in many prior DMAs was lack of axisymmetry of the entering aerosol. An interesting practical question is if the methods previously proposed to avoid this inlet asymmetry would have been as effective as our perforated ring. The little benefit of changing the slit width found previously as well as here suggests strongly that viscous pressure drop along the perimeter of the slit (as hypothesized by Eichler) is not the main cause of the asymmetry observed. A more likely source of asymmetry is the fact that the aerosol inlet flow enters the annular chamber as a jet of considerable speed, which needs to be redirected tangentially. This process requires the formation of a stagnation point in the vicinity of the aerosol inlet tube, bringing the local pressure to its stagnation value, which is considerably higher than the static pressure in other regions of the annular chamber. For the thinnest slit used here, its estimated exit area of 5.5 mm$^2$ is comparable to the area of the 3 mm inlet tube (7.06 mm$^2$). The velocities of the round jet entering the annular chamber and the annular jet leaving it are hence also comparable. Therefore, the pressure drop across the slit is comparable to the excess pressure in the tube inlet region, resulting in substantial lack of symmetry of the flow through the slit. This problem cannot be solved by Eichler's strategy. It is reasonably well (though not perfectly) solved by the perforated ring. With the ring in place, the incoming round aerosol jet still stagnates against the entry face of the ring, resulting in larger than average aerosol flow through one hole facing the round entering jet. However, the pressure in the region between the exit face of the ring and the slit is relatively uniform, except perhaps at the point where the jets formed by the ring holes stagnate against chamber surfaces. The effect of such local asymmetries is evidently much weaker than that of the original aerosol jet, given that the corresponding flow rate has been divided by the (large) number of holes.

Reischl's tangential inlet would at first sight seem to be able to avoid the formation of a stagnation point, perhaps precluding the undesirable local pressure rise just discussed. The matter is however not so clear, as the impingement of an inviscid jet into a wall results in a stagnation point, even when the incoming jet forms a small angle with the tangent to the wall. Even in the case of a real viscous flow, it is evident that the pressure distribution in the annular chamber must be periodic in the polar angle $\theta$. A point of maximal pressure must therefore exist at a certain $\theta$, and the flow around the annulus must change direction at this point. In other words, it is not possible that the entering jet would make a full turn around the annulus because the pressure at the point where the aerosol enters the annulus cannot have the maximal and minimal pressure at the same time. Evidently, if the flow moves clockwise in a certain region of the annulus and counter clockwise in another, it must stagnate somewhere in between, very much as the inviscid impinging jet. These general considerations evidently do not preclude the possibility of a successful tangential inlet scheme, but its effectiveness would need to be demonstrated.

Once the problem associated to the formation of stagnation regions somewhere along the slit circumference is identified, other remedies to the resulting inlet asymmetry problem may be identified. One possibility is to widen the area of the inlet tube, though at the cost of increasing the response time and the space charge losses (particularly serious in the case of naturally charged electrospray sources). Another possible solution is to place a deflector surface facing the inlet aerosol flow, such that the sample is symmetrically divided into two tangential streams, one moving clockwise, another counterclockwise. A well-designed deflector would create a sufficiently uniform velocity distribution on the two sides of the entry region to the annular chamber, minimizing local regions of high static pressure in the vicinity of the slit.

**Pending improvements**

The more symmetric sample inlet flow obtained with perforated rings has pushed to larger sample flow rates the flow instabilities typical of the mixing region. However, these instabilities remain at high enough *q*, and do still limit the flexibility of the instrument. Other more effective approaches to further delay such instabilities are not only most desirable, but surely also possible.

Another important limitation of the present study is in the quality of the protein size standard used, precluding the determination of resolving powers larger than 25-28.

**5. Conclusions**

1) A new method to evaluate (and improve) DMA performance with particles considerably larger than existing monomobile molecular standards has been developed. It relies on the previously known use of charge-reduced electrosprayed proteins, whose finite peak width may be precisely determined with existing nanoDMAs. Convolution of this known size distribution with the Knutson-Whitby triangular transfer function for non-diffusing particles provides an ideal baseline to identify and correct flow problems in a wide range of sheath gas and sample flow rates that had not been previously studied. The finite mobility distribution of the protein aerosol, however, limits to 25-28 the maximum measurable resolving power.

2) The new testing method has been used to characterize the Halfmini DMA at the small sheath gas flow rates required to classify particles close to its ideal maximum size limit of 30-40 nm. The original instrument is seen to respond poorly under such low flow rates, as increasing the ratio $q'$ of sample to sheath gas decreases resolution far more than theoretically expected. Non-ideal behavior sets in after exceeding a certain critical value $q'^*$, which depends on the sheath gas flow rate $Q$.

3) Largely removing the radial component of the annular aerosol stream as it joins the sheath gas is ineffective in removing flow non-idealities, as is changing the slit width.

4) The axial symmetry of the aerosol flow as it passes from the annular chamber into the inlet slit is greatly improved by introducing a perforated ring upstream from the slit. As a result of this measure, $q'^*(Q)$ increases substantially. This widens the domain of ($q, Q$) offering nearly ideal DMA performance, now including the flow rate range required to cover particle diameters of 30 nm and beyond. The range of $q$ values resulting in high resolution is also considerably widened at large sheath gas flows. This enables high-resolution studies of clusters at unusually large sample flow rates, with great advantage in the sensitivity with which such investigations may be undertaken.

5) The main cause of the original asymmetry in the annular aerosol flow is traced to the high velocity of the sample gas as in enters through a single tube of small bore into the annular chamber preceding the inlet slit. This results in stagnation regions near the slit region closest to this sample inlet jet, where the static pressure is substantially larger than in other regions of the slit circumference. While this problem is well solved by our perforated rings, we believe it would not be solved by methods previously proposed. Alternative solutions are nonetheless available, including the use of a wider inlet tube, or a deflector surface that would turn by 90$^o$ the entering aerosol flow into two opposed streams moving tangentially along the walls of the annular chamber.

6) Flow instabilities associated to increasing $q/Q$ have been pushed to larger $q/Q$ values by the perforated ring. However, instabilities persist in other regions of ($q, Q$) space, limiting the flexibility of the instrument.

7) The principles successfully tested with the Halfmini DMA apply widely to other DMAs.

**Acknowledgment**.
This study has been supported in part by SEADM through *Technologies for low emission light duty powertrains* (GV-02-2016), Grant Agreement Number: 724136. The author is most grateful to Mr. Luis J. Perez-Lorenzo for his key original contributions to the concept of a *circularizer ring*. We thank Dr. Athanasios Konstandopoulos for pointing out the need to extend the upper size range of the Halfmini DMA to permit comparison of new nanoparticle measurement techniques with already established techniques for larger particle characterization in automobile exhausts. This project has received considerable intellectual stimulus from a collaboration with Mr. Jerome J. Schmitt of NanoEngineering Corporation, aimed at high resolution size classification of viral particles 20-60 nm in diameter.

**Conflict of interest statement**. The author declares a personal interest in the company SEADM commercializing the DMA and the electrometer used in this work.